\documentclass[journal=jpclcd,manuscript=article,layout=twocolumn]{achemso}

\usepackage[version=3]{mhchem} 
\usepackage{graphicx}
\usepackage{physics}
\usepackage{amssymb}
\usepackage{subcaption}
\usepackage{placeins}
\usepackage{siunitx}
\usepackage{textcomp}
\usepackage{gensymb}
\usepackage{color}

\newcommand{\rev}[1]{\textcolor{black}{#1}}
\newcommand{\Rorder}[1]{\textcolor{black}{#1}}

\author{Nakul Pande}
\email{n.pande@utwente.nl}
\affiliation[University of Twente]
{Physics of Fluids, University of Twente, Enschede}
\alsoaffiliation{Photo Catalytic Synthesis, University of Twente, Enschede}

\author{Shri K. Chandrasekar}
\affiliation[University of Twente]
{Photo Catalytic Synthesis, University of Twente, Enschede}

\author{Detlef Lohse}
\affiliation[University of Twente]
{Physics of Fluids, University of Twente, Enschede}

\author{Guido Mul}
\affiliation[University of Twente]
{Photo Catalytic Synthesis, University of Twente, Enschede}

\author{Jeffery A. Wood}
\affiliation[University of Twente]
{Soft Matter, Fluidics and Interfaces, University of Twente, Enschede}

\author{Bastian T. Mei}
\email{b.t.mei@utwente.nl}
\affiliation[University of Twente]
{Photo Catalytic Synthesis, University of Twente, Enschede}
\author{Dominik Krug}
\email{d.j.krug@utwente.nl}
\affiliation[University of Twente]
{Physics of Fluids, University of Twente, Enschede}

\title[\texttt{achemso} 1D pH profiles]
{Electrochemically Induced pH Change: Time-Resolved Confocal Fluorescence Microscopy Measurements and Comparison with Numerical Model}

\begin{document}
\begin{tocentry}
\includegraphics{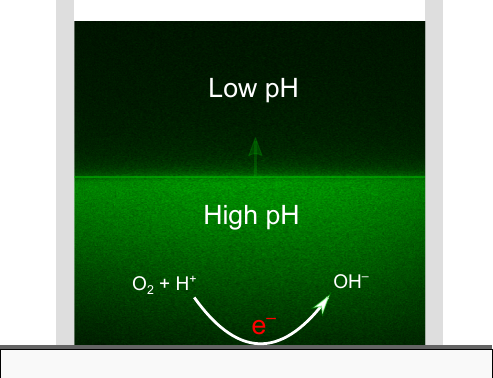}





\end{tocentry}

\begin{abstract}

Confocal fluorescence microscopy is a proven technique, which can image near-electrode pH changes. For a complete understanding of electrode processes, time-resolved measurements are required, which have not yet been provided. Here we present the first measurements of time-resolved pH profiles with confocal fluorescence microscopy. The experimental results compare favorably with a one-dimensional reaction-diffusion model; this holds up to the point where the measurements reveal three-dimensionality in the pH distribution. Specific factors affecting the pH measurement such as attenuation of light and the role of dye migration are also discussed in detail. The method is further applied to reveal the buffer effects observed in sulfate-containing electrolytes. The work presented here is paving the way toward the use of confocal fluorescence microscopy in the measurement of 3D time-resolved pH changes in numerous electrochemical settings, for example in the vicinity of bubbles.

\end{abstract}

Electrochemical reactions in aqueous solutions are strongly affected by the pH near the electrode. In corrosion science, potential-pH phase diagrams\cite{Pourbaix1974} (Pourbaix diagrams) best summarize this relationship. Moreover, in applications of energy storage and material conversion (e.g., \ce{CO2} and \ce{N2} reduction to useful products), where protons in solution are consumed, there is a direct link between the pH and the efficiency of the electrochemical cell. Measuring and understanding pH profiles near electrodes is therefore essential and can provide insight into the local surface chemistry and help design efficient electrochemical systems. This is particularly relevant in the reduction of \ce{CO2}, where sensitivity to the near electrode pH may limit the desired product formation. \cite{Kas2015,Yang2019,Burdyny2019} 

An effective technique to detect pH changes is the use of indicator molecules, such as  fluorescein, whose fluorescence changes with pH. Unlike point measurements, e.g., via scanning electrochemical microscopy,\cite{Monteiro2020} imaging fluorescence fields allows for spatially resolved pH-measurement. When coupled with confocal microscopy, this approach offers an even higher spatial resolution, and has already demonstrated its potential in electrochemical applications.\cite{Bouffier2017} For example, Unwin et al.\cite{Cannan2002,Rudd} measured three-dimensional steady-state pH profiles on microelectrodes. \citeauthor{Cannan2002}\cite{Cannan2002} determined the pH change accompanied by the reduction of benzoquinone to hydroquinone. Similarly, \citeauthor{Rudd}\cite{Rudd} measured the pH profiles induced by the reduction of water and oxygen on gold electrodes. They considered different electrode shapes and compared their results with a steady-state reaction-diffusion model. \citeauthor{Leenheer2012}\cite{Leenheer2012} applied the fluorescence method in a flow cell to compare the steady-state pH profiles formed (for hydrogen evolution) on patterned \ce{Au} electrode surfaces. Furthermore, they measured pH profiles on various electrode materials, thereby suggesting this technique as a screening tool for identifying electrocatalysts. Similar to \citeauthor{Rudd},\cite{Rudd} they compared their measurements with a steady-state model, one including laminar flow.

Although fluorescent measurements of spatiotemporal pH profiles near ion-selective membranes have been recently undertaken,\cite{Mai2012a} with related electrokinetic modeling by \citeauthor{Andersen2014a},\cite{Andersen2014a} such measurements are lacking for electrolytic systems and near the electrodes. Here, besides electric field effects, large gradients in pH are created because of chemical reactions at the electrode surface. Time resolution is then essential to capture the dynamics at the electrode-electrolyte interface. One such application would be the measurement of pH profiles around growing hydrogen bubbles in solution, which may reveal transient reaction hot-spots.\cite{Pande2019,Angulo2020} Similarly, other situations involving phase change, simultaneous electrode reactions, or bulk buffer reactions during electrolysis require time-resolved measurements for their accurate characterization. Certainly, a further development of time-resolved measurements techniques is urgently needed to understand dynamic processes occurring at electrode/electrolyte interfaces in electrochemical processes. In spite of the need, to the best of our knowledge, a quantitative comparison of time-resolved pH measurements and modeling using fluorescent dyes is not yet available in the literature. \Rorder{In this contribution we demonstrate the feasibility of using fluorescent dyes to measure spatiotemporally varying pH profiles in solution by comparing the pH changes arising from electrochemical oxygen reduction with a time dependent reaction-diffusion model. We further apply this technique to highlight buffer effects in sulfate-containing electrolytes.}

\subsection{Confocal fluorescence (pH) microscopy}

\begin{figure}
\includegraphics[width=0.8\columnwidth]{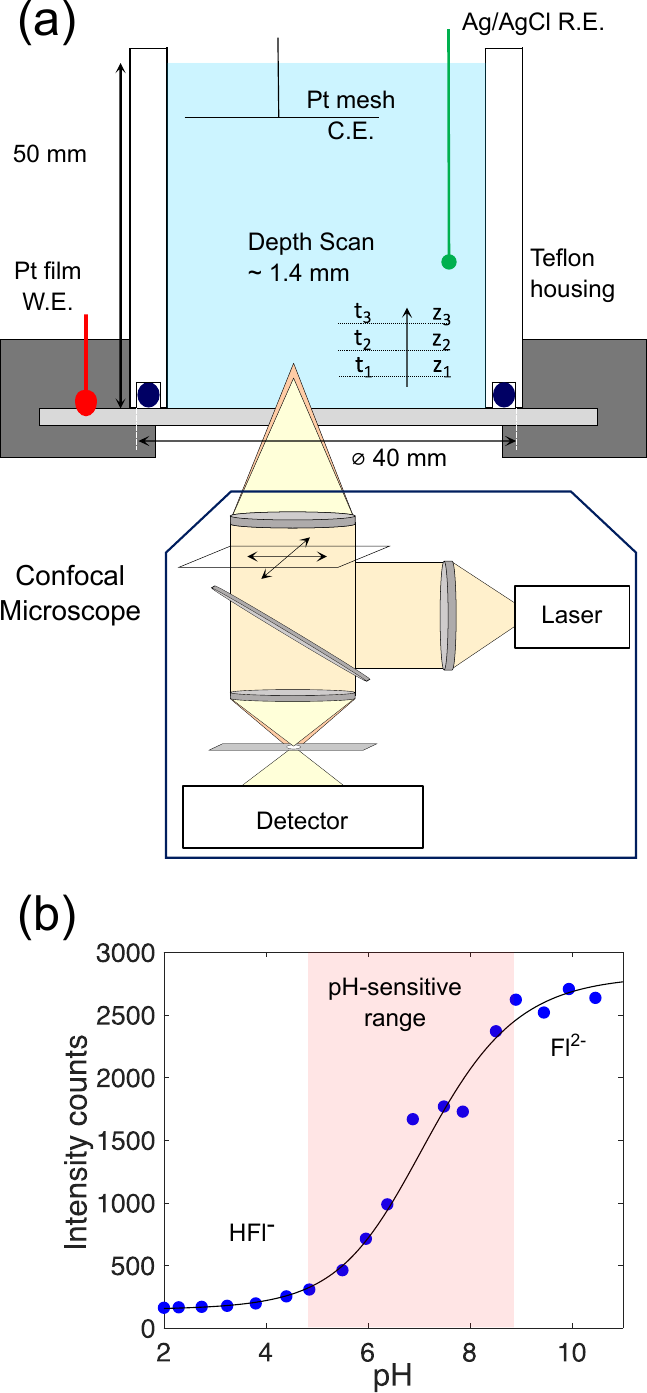} 
\caption{(a) Schematic of the experimental setup. The electrochemical cell was placed on the inverted confocal laser scanning microscope. The transparent working electrode allowed for depth-wise ($z$) measurement of fluorescent intensity. (b) Calibration results for the  pH dependence of fluorescein. The experimental data (filled circles) shown here are the mean of three measurements of intensity measurements at each pH (error bars are smaller than the marker size). A sigmoidal function (black line; see Supporting Information for details of fit) is fitted to all three measurements at each pH. The measurements were performed in 1 mM \ce{Na2SO4} containing 8$\mu\text{M}$ fluorescein. pH was adjusted to the required value by addition of \ce{H2SO4}. }
\label{fig:schematic}
\end{figure}
\Rorder{
To carry out the measurements, an electrochemical cell assembly was mounted on top of an inverted confocal microscope. A schematic of the setup which also containts the relevant dimensions is shown in Figure \ref{fig:schematic}a. The electrochemical housing was made of Teflon. In all measurements, a platinized titanium mesh was rolled up and placed as a ring at about 4 cm from the working electrode. This assembly successfully prevented any interference of the counter electrode reaction with the pH measurement. A 10 nm thick platinum film evaporated on a circular glass slide (thickness 170 $\mu$m, diameter 50 mm),  formed the working electrode. A BASi\textsuperscript{\textregistered} Ag/\ce{AgCl} (in 3M NaCl) was used as a reference electrode. Unless otherwise stated, 0.5M \ce{NaClO4} with 8 $\mu$M Sodium Fluorescein was used as electrolyte. The measurements involved taking fluorescence images along the scanned direction $z$ (in a serial fashion as shown in Figure \ref{fig:schematic}a). The mean of each image was then taken as the measured intensity at the corresponding $z$-position. The fluorescence signal was found to be compromised up to a distance $z \approx $ 100 $\mu$m above the surface (see experimental section and supporting information for further details). Therefore, any fluorescence intensity (and resulting pH) information was only obtained above this threshold. }

\Rorder{Fluorescein (Fl) is a popular choice to probe pH changes in electrochemical cells. \cite{Bowyer1996,Cannan2002,Rudd,Leenheer2012,Tassy2020} The pH sensitivity of Fl} \rev{is well documented \cite{Diehl1985,Diehl1989b,Sjoback1995} and} \Rorder{arises from the existence of different protonated forms of the molecule in solution.} \rev{The fluorescence emission of constant pH solutions was measured for the intensity-to-pH calibration (as shown in Figure \ref{fig:schematic}b)}. \Rorder{These results were fitted with an analytical function to allow for a conversion from  Fl intensities obtained in the experiments to pH. The laser and confocal settings were kept  constant throughout the study such that the curve in Figure \ref{fig:schematic}b applies to all experiments. The dye is found to be particularly pH-sensitive in the range $5 \lessapprox pH \lessapprox 10$ (indicated by the shading in Figure \ref{fig:schematic}b), as evidenced by the pronounced increase of fluorescence emission intensity measured with increasing pH within this interval. However, the highly nonlinear intensity-pH relationship suggests that measurement results beyond $pH \gtrapprox 8.5$ value should be taken with caution. Details on the fit and the repeatability are provided in the supporting information.} \rev{The effect of dye migration induced by the electric field has also been addressed therein. Migration is detrimental to the present technique, which assumes a homogeneous dye distribution. It was found that a certain minimum supporting electrolyte concentration is necessary to keep migration effects at bay}.

\Rorder{Last, to compare with experimental results, we also simulated the pH profiles during reaction. Since the supporting electrolyte concentration is high, we adopted a reaction-diffusion model for the simulations of the general form
 \begin{equation}
    \pdv{c_k}{t} = D_k\pdv[2]{c_k}{z} \pm f(c),
    \label{eq:1}
 \end{equation}
where $c_k(z,t)$ is the concentration of species $k$, $D_k$ the corresponding diffusion constant, and $f(c)$ is a non-linear function representing reaction terms. Here the chemical reactions considered are
\begin{align}
    \ce{  H+ + OH- &<=>[k_f][k_b] H2O}\\
    \ce{  H+ + Fl^2- &<=>[k_{f_{Fl}}][k_{b_{Fl}}] HFl-},
\end{align}
where the equilibrium constants are  $\frac{k_b}{k_f} = K_W$ and $\frac{k_{b,{Fl}}}{k_{f,{Fl}}} = K_{f,{eq}}$. The p$K_a$ of \ce{H2Fl} is lower than the pH considered here and therefore it can be ignored. Assuming that the concentration of water is large and therefore essentially constant during the experiment, the equations have been solved for the concentration of three species: \ce{H^+}, \ce{OH^-}, \ce{Fl^{2-}}. Further details of the exact reaction-diffusion equations (with their boundary conditions) and the numerical technique (including validation) are presented in the supporting information. The contribution of capacitive current has also been taken into account. However, since the exact value of the capacitance (C) is not known in our measurements, results for C values in the range $0 \leq C \leq 120~ \mu\text{F}/\text{cm}^2$ have also been presented in Figure \ref{fig:ExpSimComp}c. Additional details can be found in the supporting information.} 

\begin{figure}
\includegraphics[width = \columnwidth]{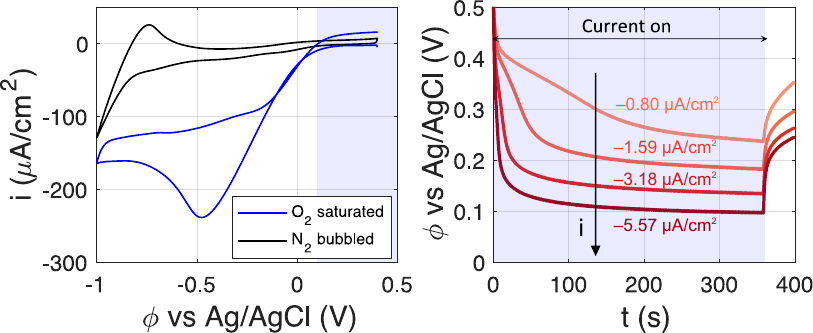}
    \caption{Left: First cycle of the cyclic voltammogram (CV) measured at 10 mV/s for the \rev{Pt working electrode in} \ce{O2} saturated and \ce{N2} bubbled solutions \rev{(pH 5 \ce{HClO4} + 0.5M \ce{NaClO4} + 8$\mu$M Fl)} in our setup. The shaded region shows the potential range measured in chronopotentiometric (CP) experiments. Right: The CP curves obtained for the \ce{O2} saturated case. The shaded region indicates the time over which the constant current is applied. The corresponding current density for each curve is mentioned as well.}
    \label{fig:CVandPotential}
\end{figure}

\subsection{Time-resolved pH measurements}
A cyclic voltammogramm (CV) of an \ce{O2} saturated solution \rev{(pH 5 \ce{HClO4} + 0.5M \ce{NaClO4} + 8$\mu$M Fl)}, along with the measured potential \rev{of the Pt working electrode} for some of the constant current experiments is shown in Figure \ref{fig:CVandPotential}. For reference, an additional CV is included for the same configuration but with a solution bubbled with \ce{N2}. \Rorder{For the current densities, $i$, and run-times considered here, it is estimated (using an initial concentration corresponding to 1 atm \ce{O2} pressure) that the oxygen at the electrode surface never gets completely depleted. This is also evident from the differences between the CV's of the \ce{O2} saturated and \ce{N2} bubbled solutions in Figure \ref{fig:CVandPotential}. Hence it can be concluded that oxygen reduction, and not water or proton reduction, is the primary reaction occurring at the electrode}. The measured potential window of operation in our constant current experiments is between 0.1V and 0.5V vs Ag/AgCl (at a starting pH of 5), which translates to 0.6V - 1.1V vs RHE. \rev{Given that an ovepotential of \cite{Li2013} $\abs{\Delta\phi} \gtrapprox$ 0.3V (over the thermodynamic potential of 1.23 V vs RHE) is required to drive the \ce{O2} reduction reaction on platinum, the potentials measured in our experiments are consistent with \ce{O2} reduction occurring at the electrode}. \rev{However, since the \ce{O2} reduction reaction depends on the pH of the solution,\cite{Si2014,Pletcher1993} the proper flux boundary conditions for \ce{OH-} and \ce{H+} are complicated (consumption of \ce{H+} or production \ce{OH-} depending on the reaction). Nevertheless, the calculated pH profiles shown here were found to be independent of this}.
Finally, it is important to mention that fluorescein is stable under the conditions applied. \cite{Compton1985,Compton1987a,Compton1988a} 

\begin{figure*}[ht]
\includegraphics[width = \textwidth]{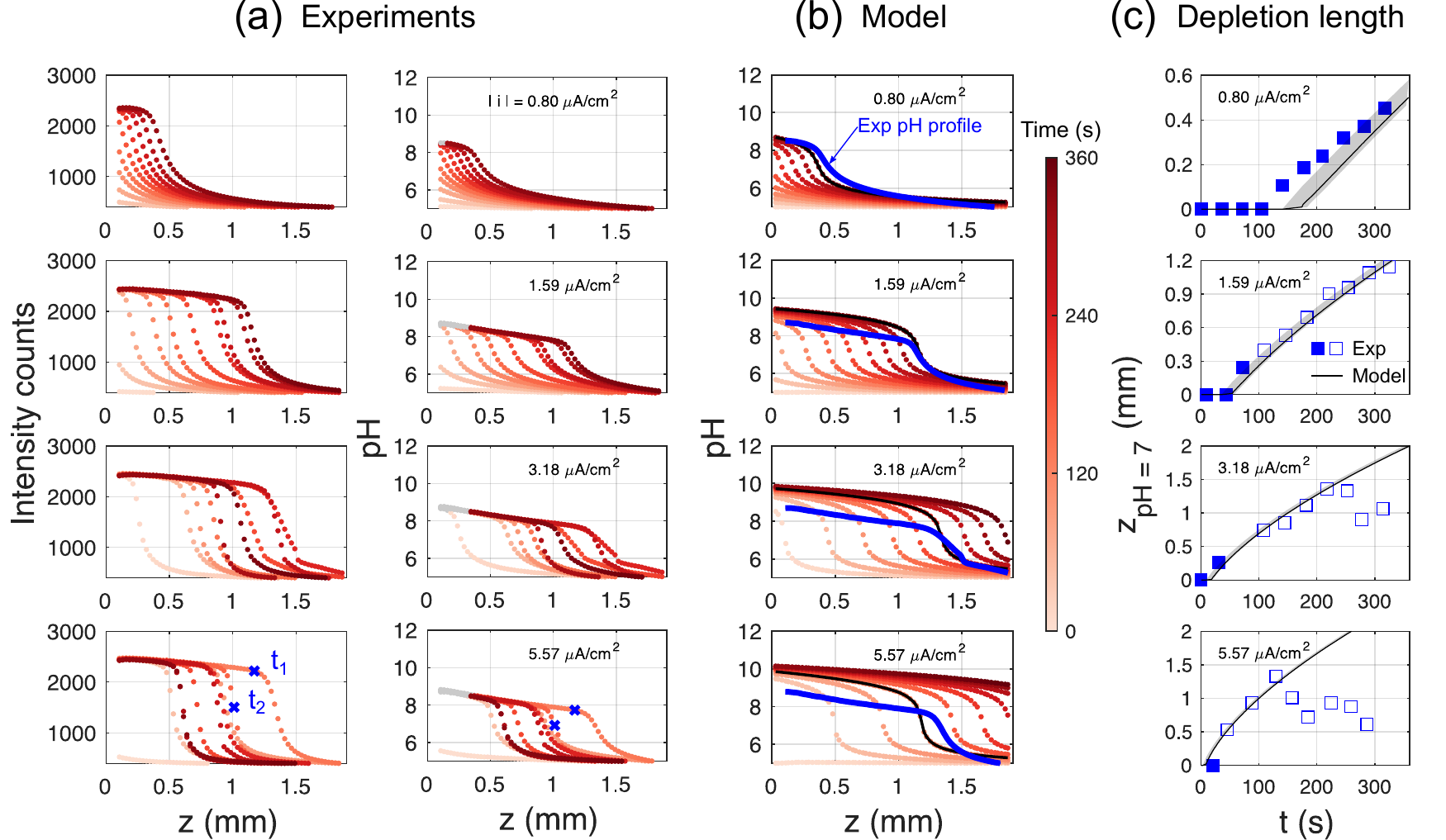} 
    \caption{Experimental versus model results. The experimental measurement is restricted to $z>0.1$ mm because of limitations of the optical setup (see supporting information for details) (a) Attenuation and depth corrected fluorescein intensity profiles. Corresponding pH profiles calculated using the calibration curve. pH $>$ 8.5 has been grayed out because of the uncertainty in measurement described in the text. $t_1$ and $t_2$ have been marked for use in Figure \ref{fig:onsT}. (b) Model results (with C = $88~ \mu\text{F}/\text{cm}^2$) calculated at experimental times. The solid lines are drawn to compare pH profiles of the experiment (blue) with the model (black). (c) Comparison of depletion length $z_{pH ~ 7}$ of experiment versus model. The shaded region indicates the model results over a range of capacitance $0 \leq C \leq 120~ \mu\text{F}/\text{cm}^2$ (solid line with C = $88~ \mu\text{F}/\text{cm}^2$). Open squares show the location of the pH front after the first appearance of the inhomogeneity as shown in Figure \ref{fig:onsT}.}
    \label{fig:ExpSimComp}
\end{figure*}

The obtained emission intensity profiles and the resulting pH distributions (at \rev{0.5 M \ce{NaClO4}} supporting electrolyte concentration) are summarized in Figure \ref{fig:ExpSimComp}a for various \rev{(constant) applied} current densities. In all cases considered here, $i$ is limited to values traditionally considered minute for electrochemistry. Despite such low current densities, the pH change and the corresponding thickness of the depletion layer are significant. 
Figure \ref{fig:ExpSimComp}a shows the attenuation corrected mean intensity of fluorescein emission as a function of distance $z$ from the electrode surface. 
Independent of the applied current density, a steep front is seen to propagate into the solution already at early times, t $<$ 300 s. This feature also translates to the corresponding pH-profiles.  It should be noted, however, that intensity levels within the resulting `shoulder' close to the electrode reach the saturation limit and due to the uncertainties described above, pH-results are grayed out in these instances. Nevertheless, the experimental results are in good agreement with the simulated pH profiles shown in Figure \ref{fig:ExpSimComp}b (see solid lines). Interestingly, also in the simulations the pH is near constant close to the electrode for $\abs{i} \geq$ 1.59 \rev{$\mu\text{A}/\text{cm}^2$}, yet with $pH = 9-10$ the values are slightly outside the experimental sensitivity range. Even at current densities of $\sim$ 1 \rev{$\mu\text{A}/\text{cm}^2$}, the depletion layer or the penetration depth of the pH profile reaches $\sim$ 1 mm into the electrolyte. At higher current densities, this depletion length grows faster and extends further into the bulk of the solution. 
\begin{figure}
\centering
  \includegraphics[width=\columnwidth]{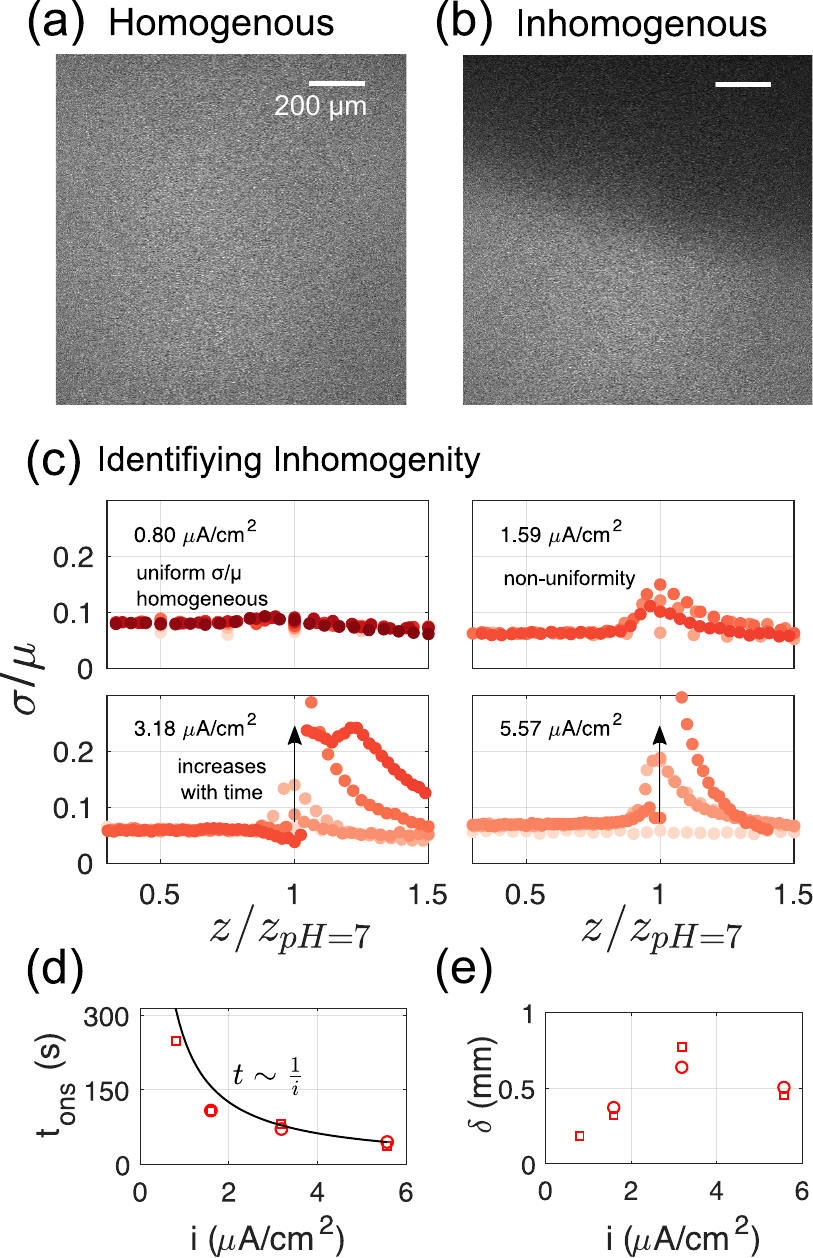}  
  \caption{Inhomogeneous fluorescein intensity in a plane. Example of fluorescein intensity images ($\abs{i}$= 5.59 $\mu A/cm^2$) at times marked in \ref{fig:ExpSimComp} a: (a) homogeneous image at time $t_1$, and (b) inhomogeneous at time $t_2$. (c) $\sigma/\mu$ versus depth for all current densities. The same color bar as in Figure \ref{fig:ExpSimComp} applies. Sharp changes are used to pick out times and positions where this nonuniformity is observed. (d) The onset time $t_{ons}$ and (e) the location $\delta$ at which the inhomogeneity is first observed vs current density. Different symbols are repeat measurements. The solid line in panel d corresponds to $t_{ons} = \frac{250}{i}$ s and is arbitrarily chosen to highlight the inverse relationship between $t_{ons}$ and $i$.}
  \label{fig:onsT}
\end{figure}

It can be seen, however, that for the two highest current densities considered here, the intensity as well as the pH profiles recede at later times (corresponding to darker shadings of the markers), whereas the model predicts a monotonic outward propagation of the front. To enable a quantitative comparison, we track the  position $z_{pH~7}$ at which the pH = 7 is encountered as a proxy for the front location. As Figure \ref{fig:ExpSimComp}c shows, the pH front propagation in the experiments is well captured by the model for the two lower current density cases presented here. At higher current densities and at late times, though, the pH front in experiments either recedes or saturates. This is also true for repeat measurements made (see supporting information). However, this effect appears to be an artifact of the way the mean fluorescein intensities are calculated. Consistent with the 1D assumption, only a measure of the mean across the entire image (i.e. a plane parallel to the electrode) is considered. For example, at $t_1$ (for $\abs{i} =$ 5.59 $\mu A/cm^2$, see Figure \ref{fig:ExpSimComp}a), this is appropriate as highlighted in Figure \ref{fig:onsT}a. At $t_2$ though, the intensity distribution displayed in Figure \ref{fig:onsT}a becomes distinctly inhomogeneous as seen in Figure \ref{fig:onsT}b. This implies that 2D or 3D effects become relevant, which are not captured in the one-dimensional model.

To determine the location and time at which 3D effects become relevant, we consider the  standard deviation ($\sigma$) normalized with the mean intensity ($\mu$) of the image as shown in Figure \ref{fig:onsT}c. To minimize the effect of high frequency spatial noise, the image was box-filtered with a filter size of 50 pixels before calculating $\sigma$. Figure \ref{fig:onsT}c captures the uniform image intensity for $\abs{i}$ = 0.8 \rev{$\mu\text{A}/\text{cm}^2$} as a near constant $\sigma/\mu$. In contrast, a visible peak in $\sigma/\mu$ at the depletion front $z = z_{pH~7}$ is observed for all other cases. At the two highest current densities considered, the unsteadiness in fluorescein intensity develops over time as well. The onset time ($t_{ons}$ defined as $\sigma/\mu > 0.1$) of this instability thus calculated is, in Figure \ref{fig:onsT}d, found to sharply reduce with increasing current densities which is well approximated by an inverse proportionality. It is conceivable then that this instability occurs only after a certain threshold number of \ce{H+} ions have been depleted from solution. The distance $\delta$ at which this non-uniformity is first measured, shows no clear trend: the non-uniformity first increases until $\abs{i}$ = 3.18 \rev{$\mu\text{A}/\text{cm}^2$} and then decreases again slightly later. Since we only look at a small portion of the electrode though, deviations from a 1-D profile can occur much earlier, at a different $\delta$. It is unlikely that the reaction at the counter electrode plays any role in the appearance of instability as it is sufficiently far away compared to the measured depletion lengths of $\sim$ 2 mm. Possible reasons could then be the presence of electric field effects or induced fluid flow in the system \cite{Mani2020}, which have not been modelled. However, despite the early appearance of inhomogeneity, the pH profiles in experiments are similar to the model results up to distances and times that are much larger (see Figure \ref{fig:ExpSimComp}c: filled and open symbols, and blue/black lines in \ref{fig:ExpSimComp}b). It may be possible then that the departure of pH profiles in experiments, from a 1D diffusion approximation, occurs only after a certain minimum $\sigma/\mu$ (and corresponding inhomogeneity) is reached.
\begin{figure}
\includegraphics[width=\columnwidth]{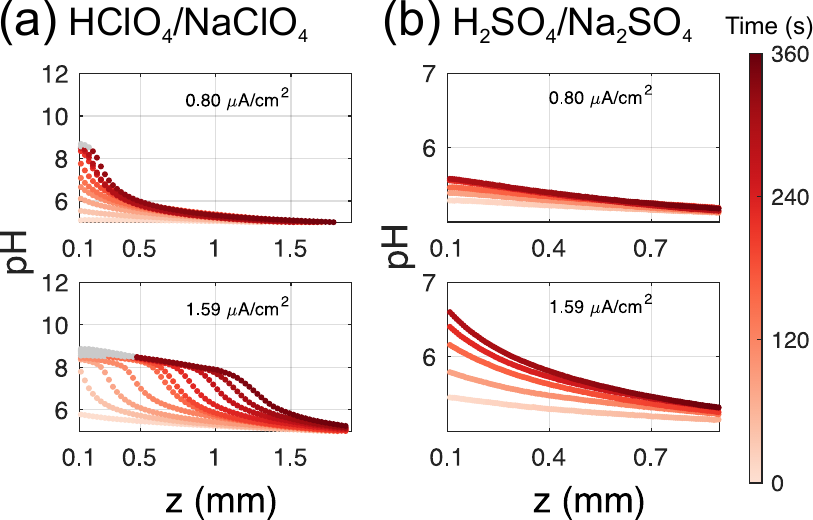}
    \caption{Comparison of pH profiles for (a) perchlorate (\ce{NaClO4}/\ce{HClO4}) and (b) sulfate (\ce{Na2SO4}/\ce{H2SO4}) electrolytes. All measurements were performed in 0.5M supporting salt concentration containing 8$\mu \text{M}$ Fl. pH was adjusted to pH 5 by addition of respective acid. All solutions were bubbled with \ce{N2} before starting the experiment. pH $>$ 8.5 has been grayed out because of the uncertainty in the measurements described in the text.} 
    \label{fig:sulphates}
\end{figure}

\subsection{Sulfate buffer effect}
In addition to the above measurements, we proceed to evaluate the developing pH profiles in sulfate containing electrolytes, e.g. in the \ce{Na2SO4}/\ce{H2SO4} system. This system is frequently used (for example sulfuric acid is commonly used to study \ce{O2} reduction) but, in contrast to perchlorate electrolytes, may induce additional buffer capacity, thus changing the pH profiles. In fact, \ce{H2SO4} has two dissociation constants, the second corresponds to the dissociation of \ce{HSO4-} with a p$K_a$ of around $2$ \cite{Hamer1934,Covington1964a,Wu1995a} . Figure \ref{fig:sulphates} compares the pH profiles measured for the sulfate case to those obtained with perchlorate electrolyte, for the two lowest current densites. It is evident that the pH profiles develop significantly slower in sulfate-containing electrolytes. For example, for $\abs{i}$ = 1.59 $\mu\text{A}/\text{cm}^2$, the pH profiles in the $\ce{Na2SO4}/\ce{H2SO4}$ system have no clear front propagating in the solution; the profiles rather become increasingly steep close to the electrode surface with time, while for the perchlorate solution depletion lengths $z_{pH~7}\approx$ 1.5 mm are achieved.

To try to further explain the experimental results, we consider the p$K_a$ of \ce{HSO4-}, which, although is well below our starting pH (pH = 5), due to the presence of the large concentration of \ce{SO4^2-} in solution, creates a reservoir of \ce{HSO4-} ions which act as a source of protons in solution and stabilizes the solution against pH changes. We attempt to capture this effect in the 1D model as our results in the supplement show. This buffer effect is most likely present in experimental measurements in literature with sulfate electrolytes.\cite{Leenheer2012,Monteiro2020} For example, 
\citeauthor{Leenheer2012}\cite{Leenheer2012} measured the pH on patterned gold electrodes in \ce{Na2SO4} solutions, with different pattern shapes and area. However, their steady state simulations predicted a depletion zone much larger than experiments. Similarly, the buffering effect of \ce{Li2SO4} solutions may also be present in the recent measurements by \citeauthor{Monteiro2020}.\cite{Monteiro2020} A comparison such as ours, between perchlorate and sulfate electrolytes, should help to quantify the magnitude of this effect and help better interpret results.

To summarize, we have successfully demonstrated the use of fluorescein to measure time-resolved pH profiles in solution. The results of a time-dependent reaction-diffusion model compare reasonably well with the experimental data. However, the inhomogeneity of pH in a plane that develops at `high currents' clearly shows the need for time-varying local pH measurements. The crucial aspects to consider when using fluorescence microscopy for pH measurement, like optical distortions and signal attenuation, have been carefully examined. Furthermore, the concentration of the supporting electrolyte is shown to influence migration of fluorescent dyes and should be considered to avoid pitfalls in pH measurement in electrochemical systems. For sulfate containing electrolytes, our analysis reveals buffering effects, which likely explain the difference between the measured diffusion profiles and those observed in experiments in the past \cite{Leenheer2012}.

Fluorescence microscopy offers time-resolved and relatively non-intrusive measurement of pH instantly over a large area.
Since the principle of measurement presented here is applicable to other fluorescent dyes with a different pH detection range, this technique can be used for a wide range of electrochemical systems to elucidate electrode dynamics. This holds in particular for \ce{CO2} reduction on gas diffusion electrodes, since the second p$K_a$ of carbonic acid lies in the fluorescein detection region. Our developed method can be directly implemented to quantify mass transport, the role of bicarbonate concentrations etc. in the electrolyte. More generally, the measurement technique presented here offers insight into the dynamics of ions in solution, important to many electrochemical systems, none more so than in electrochemical cells to unravel the role of start-stop transients. Detailed information of the pH distribution will provide a better understanding of electrode processes and aid in the overall design of electrochemical systems for eventual use in large scale electrolysis.
\subsection{Experimental Methods}

\Rorder{For the working electrode, a 3 nm thick chromium (under) layer was used for better adhesion between the platinum film and the glass slide.  The sheet resistance of the resulting thin film electrode was about 69 $\Omega$. The electrical connection to the working electrode was made with a platinized titanium point contact. Prior to the measurement, the pH of the solution was adjusted to a pH value of 5 by addition of appropriate amounts of 0.1 M \ce{HClO4} (or 0.5 M \ce{H2SO4} for sulfate electrolytes). The pH of the solution before the start of each experiment (as well as the calibration solutions shown in Figure \ref{fig:schematic}b) was measured using the Hannah Instruments Edge-pH meter that has an accuracy of $\pm$ 0.02 pH units. All chemicals were purchased from Sigma-Aldrich. Further experimental and numerical details can be found in the Supporting Information}.
\begin{suppinfo}
Details of experimental method, numerical method, repeat measurements and comparison of experiments and model for sulfates.
\end{suppinfo}
\acknowledgement
This research received funding from The Netherlands Organization for Scientific Research (NWO) in the framework of the fund New Chemical Innovations, project ELECTROGAS (731.015.204),
with financial support of Akzo Nobel Chemicals, Shell Global Solutions, Magneto Special Anodes (an Evoqua Brand), and Elson Technologies. We acknowledge The Netherlands Center for Multiscale Catalytic Energy Conversion (MCEC) and the Max Planck Center Twente for Complex Fluid Dynamics for financial support. D.L. also acknowledges financial support by an ERC-Advanced Grant.
\bibliography{bib/Main_Submission_Final_proofs}

\end{document}


\subsection{Experimental Methods}
\label{ExpMethod}
\begin{figure}
\centering
\includegraphics[width=\columnwidth]{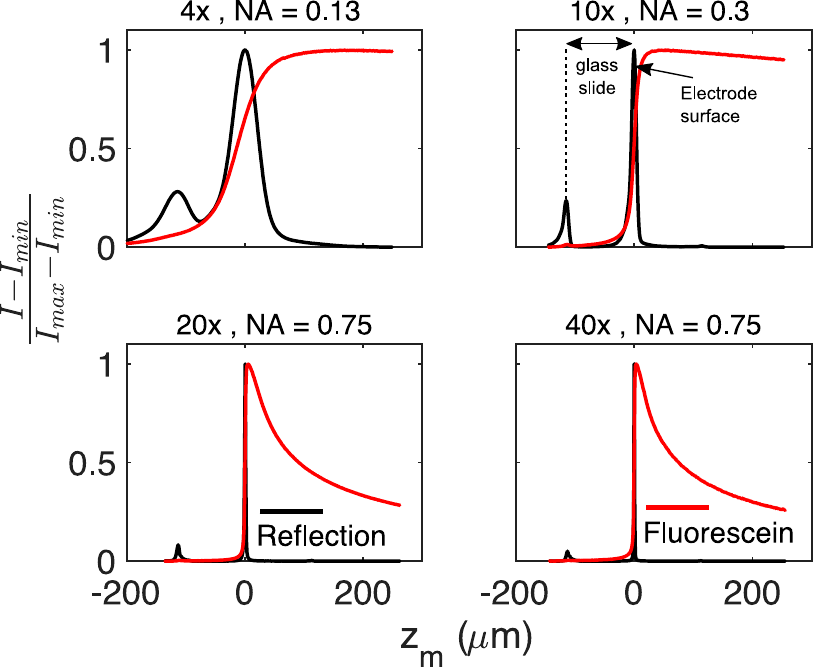}
    \caption{Surface reflection and fluorescein intensity for three different Numerical Apertures (NA's) and 4 different magnifications for a constant pH and Fluorescein concentration solution (E-cell off). Maxima in the surface reflection corresponds to the electrode surface.}
    \label{fig:RefInt}
\end{figure}

A Nikon inverted laser scanning confocal fluorescent microscope (Nikon confocal microscope A1 system, Nikon Corporation, Tokyo, Japan) with a 10x dry objective (CFI Plan Fluor 10x/0.3, numerical aperture = 0.3, working distance = 16 mm) was used in the resonant scanning mode (33 ms per image) to measure a 1.28 mm $\times$ 1.28 mm region (512 $\times$ 512 $\text{pixel}^2$) chosen close to the center of the electrode. A 488 nm excitation laser was chosen to excite Fl, while the emission was collected in a 515-550 nm wavelength window. The pinhole (29.4 $\mu$m) cuts off any out of focus light allowing to image thin volume sections. Close to 70 measurement cross sections with 20 $\mu$m distance from each other were scanned repeatedly , resulting in a total measured depth of 1.4 mm. The scanning along $z$ proceeded from below the electrode surface into the solution and was repeated at a typical rate of 2 Hz. The acquisition frequency was limited by the movement of the stage in the z-direction.
To determine the location of the electrode surface we measure the light reflection \cite{Paddock2002} from the working (glass slide) electrode. Figure \ref{fig:RefInt} shows the mean reflection and fluorescein intensity signals measured simultaneously, starting from below the glass slide. The fluorescein signal does not provide clear information on the (electrode) surface location, hence the reflection signal is used. The presence of the glass slide causes two reflection maxima 115 $\mu$m apart, which can be used to establish the surface positions. Therefore, before starting each experiment, the reflection signal is measured (by scanning optical sections 5 $\mu$m apart) to determine where the surface of the electrode is located. It should be noted that even though a simplified Point Spread Function (PSF) has been assumed in the following analysis, the actual PSF will likely be more complicated due to the many different refractive index media (air-glass-chromium-platinum-water). While increasing the numerical aperture (NA) will increase the z-resolution\cite{Claxton2006} (z-resolution $\propto$ 1/NA$^2$), the choice was limited due to significant reduction in the measured fluorescent intensity signal at the electrode interface for high NA objectives as shown in Figure \ref{fig:RefInt}. This was likely due to the high refractive index contrast between the glass-metal interface, which at the large incident angles of a high NA objective, may cause considerable loss of transmitted light due to internal reflection.  

In the experiments, sections in the scanned direction are taken by the programmed movement of the optical stage (20 $\mu$m apart in the experiments in the main text). The distances measured ($z_m$) in this way (from the electrode surface), however, do not take into account distortion in the light path due to variations in the  refractive index (air vs. aqueous electrolyte). \citeauthor{Visser2008}\cite{Visser2008} give the relationship between the actual focal distance ($\Delta z$) to stage movement ($\Delta z_m$) as $\Delta z = \Delta z_m n$, where $n$ is the refractive index of the medium. To verify the appropriateness of this correction in our case, we consider the measured glass slide thickness of 115 $\mu \text{m}$ ($\Delta z_m$).  Using $n_{glass} = 1.5$, the corrected glass slide thickness then is $\Delta z = 115~\mu \text{m} \times n_{glass} =172.5~ \mu$m, very close to the actual value of 170 $\mu m$. We carry out a similar correction for the refractive index of the electrolyte solution (using $n_{sol} = 1.33$) such that  $z = n_{sol} z_m$. 
So while the total measured depth is $z_m^{max} \approx$ 1.4 mm, the corrected depth is  $z^{max}\approx$ 1.9 mm.

\begin{figure*}
\centering
    \includegraphics[width=\linewidth]{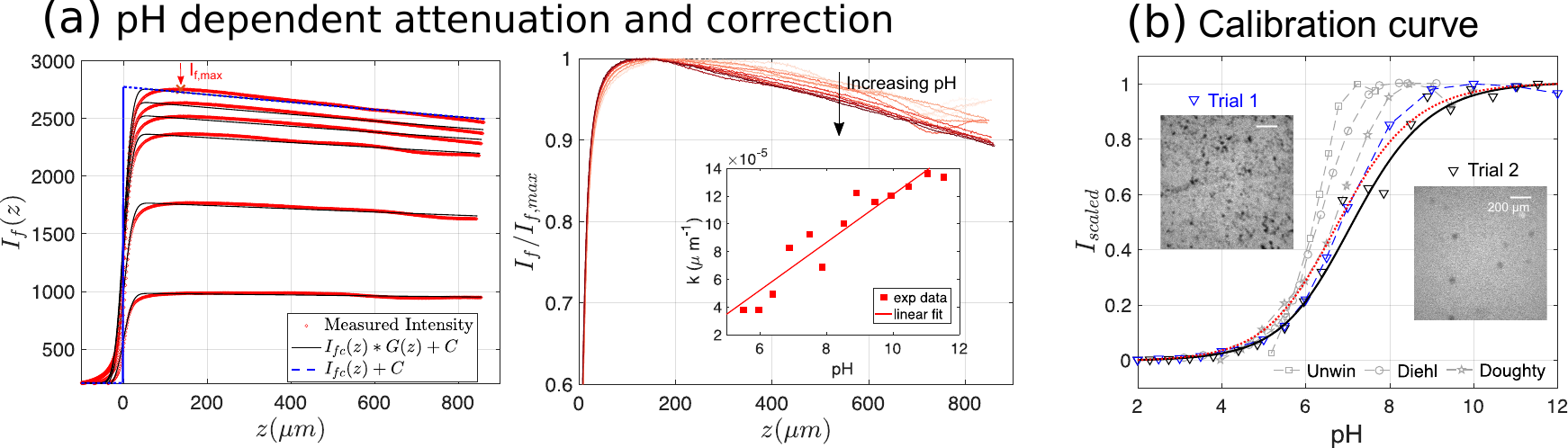} 
\caption{(a) Left: fluorescein intensity values (average over 3 runs) obtained for different pH shown by the red markers (pH $\approx$ 7.5-11.5 in steps of 1). Black line shows corresponding fit. Blue line shows the simplified fit with the obtained attenuation coefficients. Right: Normalized intensity (with their respective maximum) showing attenuation at high pH. Inset: the exponential coefficients as a function of pH. The linear fit is described by: $k = (1.7\times pH - 5.2 )\times10^{-5}~{\mu \text{m}^{-1}}$. (b) Comparison of Intensity-pH relationship of Fluorescein obtained by different authors: Unwin\cite{Cannan2002,Rudd} , Doughty\cite{Doughty2010} , Diehl\cite{Diehl1989b} against two trials measured for this study. The solid line shows the sigmoidal fit in equation (\ref{Func}), similarly scaled. The dotted red line represents the same fit adjusted to an ionic strength corresponding to 0.5 M of a monovalent salt following \citeauthor{Sjoback1995}\cite{Sjoback1995}. The dashed lines have been added for better visibility of markers.}
    \label{fig:attn}
\end{figure*}
Any measurement of fluorescence requires considering the path dependent attenuation of the excitation as well as the emitted fluorescent light. Since the numerical aperture of the objective used in our experiments is small, following \citeauthor{OHSER2020}\cite{OHSER2020}, the dependence of the emitted fluorescent light ($I_{em}(z)$) on the excitation intensity ($I_{ex}(z)$) and the concentration of fluorophore ($c(z)$) at a point $z$ in the solution, along with the fluorescence efficiency ($\alpha_1$, pH dependent in the case of Fluorescein) can be written as:
\begin{equation}
    I_{em}(z) = I_{ex}(z)\alpha_1c(z)
    \label{emit0}
\end{equation}
Since in our case, the absorbance of the fluorophore is pH dependent\cite{Sjoback1995}, and the pH itself is $z$ dependent, the path dependent attenuation of the excitation intensity can be written as:
\begin{equation}
    I_{ex}(z) = I_{ex0}\left(e^{-\int_{0}^{z}\epsilon_1(pH(\tau)) c(\tau)d\tau}\right)
    \label{excitation}
\end{equation}
where $\epsilon_1$ is the pH dependent attenuation coefficient of the excitation light and $I_{ex0}$ is the excitation intensity at $z=0$. Similarly if there is an attenuation ($\epsilon_2$) of the emitted light $I_{em}(z)$, the measured fluorescence intensity $I_f(z)$ goes as:
\begin{equation}
    I_f(z) = I_{em}(z) \left(e^{-\int_{z}^{0}\epsilon_2(pH(\tau)) c(\tau)(-d\tau)}\right)
    \label{fluor}
\end{equation}
Since we use a constant concentration of fluorophore and laser settings in all our experiments, combining equations (\ref{emit0}), (\ref{excitation}) and (\ref{fluor}):
\begin{equation}
    I_f(z) = I_{f0}\left(e^{-\int_{0}^{z}k(pH(\tau))d\tau}\right)
    \label{expStp}
\end{equation}
where $I_{f0} =\alpha_1I_{ex0}c$ is the unattenuated fluorescence intensity, and $k =(\epsilon_{1}+\epsilon_{2})c$ is the overall attenuation factor. Hence a optical path history dependent correction factor of $e^{\int_{0}^{z}k(pH(\tau))d\tau}$ must be multiplied with the fluorescence intensity, $I_f(z)$, measured at a point to get the corresponding corrected value $I_{f0}$. It should be noted that for higher numerical aperture objectives, an attenuation correction such as shown in \citeauthor{Visser2008} \cite{Visser1991} must be used.

To determine $k$, we measured the fluorescence intensity as a function of $z$ for different constant pH solutions ($k$ is constant for a constant $pH$) similar to that shown in Figure \ref{fig:attn}a. We expect the fluorescence intensity at a particular pH to be exponentially decaying step function ($I_{fc}(z)$ given by equation (\ref{expStp}) with $k = \textrm{const}$) and its maximum at the electrode surface, $z=0$. The actual profiles, however, are smooth close to the electrode surface, most likely due to the point spread function (psf in the $z$ direction) of the optical system. Taking the simplest assumption of a gaussian psf i.e. $G(z) = \frac{1}{\sigma\sqrt{2\pi}}e^{-\frac{z^2}{2\sigma^2}}$ (with the standard deviation $\sigma$), the resulting profiles must then be a convolution of $I_{fc}(z)$ with $G(z)$ , and should have the analytical form:
\begin{equation}
\begin{split}
    I_{fc}(z)*G(z)= I_{f0}&e^{-kz}e^{\frac{\sigma^2k^2}{2}}\\&(1-\frac{1}{2}erfc(\frac{z}{\sigma\sqrt{2}}-\frac{k\sigma}{\sqrt{2}})
    \end{split}
\end{equation}
where, $erfc$ is the complementary error function. Consequently, we fit a function of the form $I_{fc}(z)*G(z)+C$, with four fitting parameters $I_{f0}$, $k$, $\sigma$ and $C$ (where $C$ is a parameter related to small constant unknown effects). The resulting fit is overlaid on the original data of constant pH solutions (average of 3 measurements) in fig \ref{fig:attn}a. The fit is reasonable, however differences in the location of their maximum indicates that the psf is likely more complicated than a gaussian function. Based on the $I_{f0}$ and $k$  and $C$ obtained, we further plot $I_{fc}(z) + C$ which shows that the psf smoothening is important only at the electrode surface. In our measurements, we therefore correct only for this attenuation and not the psf smoothening. The measurement is compromised below the location of the intensity maximum ($I_{f,max}$ in Figure \ref{fig:attn}a) of the profiles ($\approx$ 100 $\mu$m) and has not been shown in the main text. The attenuation correction factor $k$ as a function of pH is plotted in the inset of figure, and has been fitted with a line. At the $n$-th stack from the electrode surface (with a distance $\Delta z = 20 \times 1.33~\mu$m), the measured fluorescein intensity then must be of the form:
\begin{equation}
    I_f(z) = I_{f0}e^{-\sum_{ \eta=0}^{n-1}k(pH(\eta))\Delta z}+C
\end{equation}
 The unattenuated fluorescein intensity ($I_{f,org}$, which is now $I_{f,org} = I_{f0} + C$) can now be calculated based on the $k$ and $C$ and is:
\begin{equation}
    I_{f,org} = (I_f(z)-C)e^{\sum_{ \eta=0}^{n-1}k(pH(\eta))\Delta z}+C
    \label{cf}
\end{equation}
Lastly, it should be noted that the correction in equation (\ref{cf}) obtained (using $k$ and $C$) is calculated at each position based on the uncorrected pH. However any error associated with this is expected to be minimal.

Figure \ref{fig:attn}b compares the intensity variation of Fluorescein emission with pH measured by the authors to corresponding results in literature. For our calibration, the maximum of the fluorescence intensity measured along the scanned direction ($I_{f,max}$ in Figure \ref{fig:attn}a), for different constant pH solutions is taken as the reference intensity. The intensity variation with pH (as shown in Figure 5b in the main text), obtained so, is rescaled to the range from 0 to 1, to render the data comparable with literature results (which were similarly rescaled where needed). The general trend matches in all cases, however, there is significant spread especially at the upper limit of the pH-sensitive region.
The different ionic strengths of the calibration solution could be a possible reason for these differences \cite{Sjoback1995}. However, this effect is small as the red dotted line, which represents our calibration curve corrected to an ionic concentration of a 0.5 M monovalent salt (using the equation in \citeauthor{Sjoback1995}\cite{Sjoback1995}; assuming that the parameter $\gamma$ in our fit behaves like an effective pKa), in Figure \ref{fig:attn}c shows. Moreover, since the fluorescence intensity at a point in the sample is related to the the local concentration of the fluorophore, the laser excitation intensity, the optical path history of the light, the voltage of the photo-multiplier tube etc., the differences could be due to a number of reasons and it is not straightforward to compare the values of fluorescence intensity, for different pH, across optical setups. Rather, laser and camera settings must remain constant between calibration and experiment for a faithful conversion of intensity values to pH. Even then, there are small differences between the repeats of our calibration measurements (new pH solutions and different electrode) in Figure \ref{fig:attn}a and these could be traced to presence of small contaminants on the electrode surface. The inset of Figure \ref{fig:attn}a shows an example of a fluorescein intensity image (at pH = 10) obtained for each of the two calibration trials done. The difference between the repeats provides a sense of the calibration error as similar (and to varying degrees) contamination could be present in the experiments. However, since the calibration curve is similar in the lower limit of the pH-sensitive region, the effect on the location of the pH shoulder is minimal. Still, trial 2 (having the cleaner electrode) is used in our measurements and is shown Figure 5b in the main text. Lastly, the intensity dependence on pH (in Figure 5b in the main text) is fit with a function of the form 
\begin{equation}
    \ln(I_{f,max})=\alpha+\frac{\beta}{1+ e^{-(pH-\gamma)}}
    \label{Func}
\end{equation}
with fit parameters $\alpha~(4.991 \pm 0.034)$, $\beta~(2.946 \pm 0.039)$ and $\gamma~(5.857 \pm 0.060)$. A robust least-squares regression with a logistic function weight (implemented as 'nlinfit' in MATLAB) is used for fitting the data. The residuals of the regression appeared to be normally distributed (at all points except the two measurement triplicates between pH 6 and 8 that are not on the line in Figure 5b).
The inverse function is therefore:
\begin{equation}
    pH_{inv}=-\ln\left(\frac{\beta}{\ln(I_{f,max})-\alpha}-1\right) + \gamma.
    \label{invFunc}
\end{equation}
\begin{figure}
\centering
    \includegraphics[width=0.8\linewidth]{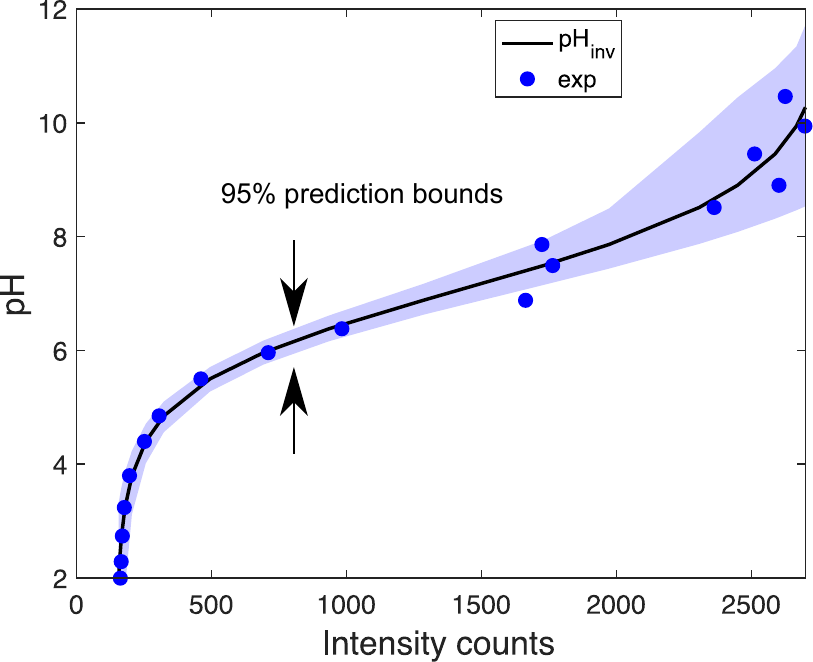}
    \caption{The inverse function based on calibration measurements presented in the main text. At a given pH, the mean of three separate intensity measurements is shown here, since the error bars are small.}
    \label{invFuncFig}
\end{figure}
The prediction interval for the non-linear fit (based on the measured calibration data) presented in Figure \ref{invFuncFig} is calculated based on \citeauthor{Guthrie}\cite{Guthrie} and shows that at the higher ends of our pH-sensitive range, the error in measurement can be close to $\pm$1 pH unit.

All remaining fits presented here use the 'fit' function in Matlab (R2019b).

\subsection{Dye migration effects: limits to supporting electrolyte concentration}
\begin{figure}
\includegraphics[width=\columnwidth]{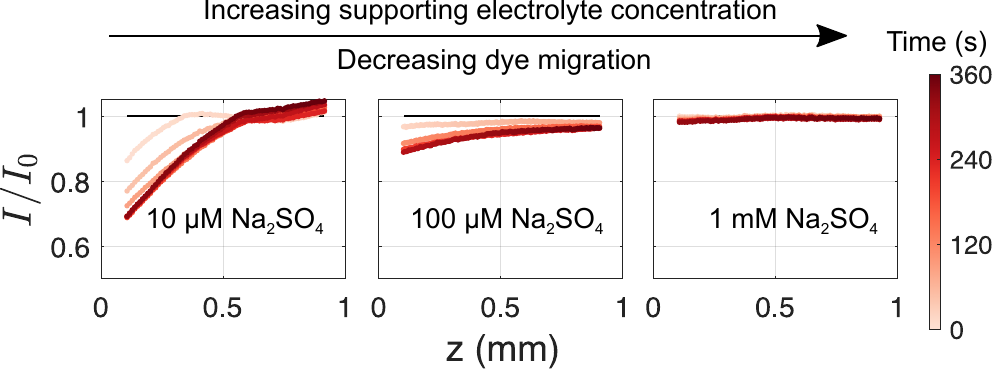}
    \caption{Migration effect for different supporting electrolyte concentrations. The emission intensity ($I$) is normalized with the intensity profile before applying the current ($I_0$).  The measurement of fluorescence intensity is restricted to $z>0.1$ mm due to limitations of the optical setup (see supporting information for details) . The measurements were performed with a solution containing 8$\mu \text{M}$ SRb. pH was adjusted to pH 5 by addition of \ce{H2SO4}. The color code reflects the time at which the concentrations were measured. $\abs{i}$ = 5.59 $\mu\text{A}/\text{cm}^2$.}
    \label{fig:figMig}
\end{figure}
Since Fl is charged in solution, an electric field induced inhomogeneity in the dye distribution will make it difficult to decouple intensity changes due to migration from a pH change.
An anionic \cite{Suzuki2016} pH insensitive dye, namely Sulforhodamine B (Sigma Aldrich, Molecular weight: 580.65 g/mol, 8$\mu \text{M}$, henceforth SRb) was therefore used to indicate the presence of dye migration. Although unlike Fl, SRb has a single negative charge in solution, due to their similar molecular weight it still serves as a good qualitative indicator of dye migration. SRb is also mildly temperature sensitive (intensity decrease $\approx 1.2\%$ per \si{\kelvin} \cite{Coppeta1998}), but for the current densities considered in this work the temperature change is estimated (even when using resistivity of pure water) to be negligible. The measured fluorescent intensity for the highest current density (5.59 $\mu\text{A}/\text{cm}^2$) in this work is presented in Figure \ref{fig:figMig}c. For low supporting electrolyte concentration, i.e. $=$ 10 $\mu \text{M}$, the fluorescence intensity of the dye near the electrode surface and up to a distance of $\approx$ 0.75 mm above the electrode decreased by up to 30 $\%$. However, this migration effect reduces significantly for increased concentration of the supporting electrolyte, and is almost negligible for \ce{Na2SO4} concentrations $\geq$ 1mM. A concentration of the supporting electrolyte much greater than this is therefore used in the experiments performed in this study.

\subsection{Details of the numerical model}
\label{model}
\begin{figure}
\centering
\includegraphics[width=0.8\columnwidth]{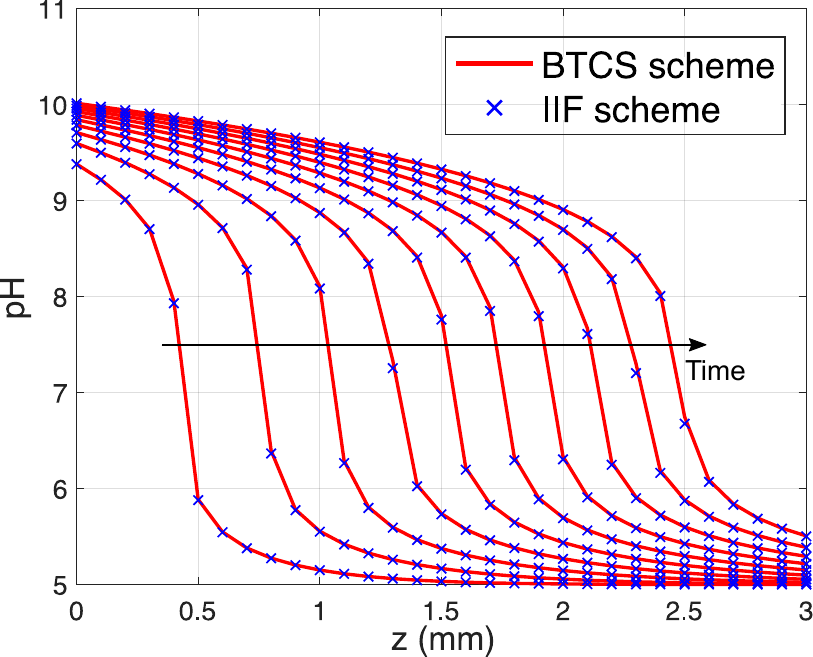}
    \caption{Comparison of results using BTCS and IIF numerical schemes. Note: only the \ce{H+} and \ce{OH-} ions were considered for this comparison. The relatively small effect of \ce{HFl-} on the pH profiles is consequently also absent.}
    \label{fig:validation}
\end{figure}
For each ion species $k$, we have a related non-dimensionalized concentration diffusion equation of the type shown in equation (1) in the main text. 
Assuming that the concentration of water is large and therefore essentially constant during the experiment the differential equations simplify to: 
    \begin{subequations}
    \begin{align}
    \begin{split}
    \frac{\partial{c_{H^+}^{*}}}{\partial{t}^{*}} &= \frac{\partial^2{c_{H^+}^{*}}}{\partial z^{*2}}-Da_1({c_{H^+}^{*}}~{c_{OH^-}^{*}}-1)\\
    &-Da_2\left({c_{H^+}^{*}}~c_{Fl^{2-}}^{*}-\frac{K_{f,{eq}}}{c_{H^+}^{0}}(1-{{c_{Fl^{2-}}^{*}}})\right)
    \end{split}\\
    \begin{split}
    \frac{\partial{c_{OH^-}^{*}}}{\partial{t}^{*}} &= D_{r,1}~\frac{\partial^2{c_{OH^-}^{*}}}{\partial z^{*2}}\\&-\frac{Da_1c_{H+}^0}{c_{OH^-}^{0}}({c_{H^+}^{*}}~{c_{OH^-}^{*}}-1)
    \end{split}\\
    \begin{split}
    \frac{\partial{c_{Fl^{2-}}^{*}}}{\partial{t}^{*}} &= D_{r,2}~\frac{\partial^2{c_{Fl^{-}}^{*}}}{\partial z^{*2}}\\&-\frac{Da_2c_{H+}^0}{T}\left({c_{H^+}^{*}}~c_{Fl^{2-}}^{*}-\frac{K_{f,{eq}}}{c_{H^+}^{0}}(1-{{c_{Fl^{2-}}^{*}}})\right)
    \end{split}
    \end{align}
    \end{subequations}
    where,
    \FloatBarrier
    \begin{table}
    \begin{tabular}{c c}
    $Da_1$ = $\frac{k_f c_{OH-}^0L^2}{D_{H^+}}$,   & $Da_2$ = $\frac{k_{f,Fl}TL^2}{D_{H^+}}$, \\
    $D_{r,1} = \frac{D_{OH^-}}{D_{H^+}}$,  & $D_{r,2} = \frac{D_{Fl^{2-}}}{D_{H^+}}$. \\
    \end{tabular}
    \end{table}
    \FloatBarrier
   
    Further $T$ is the total initial concentration of fluorescein $(c_{HFl^-}+c_{Fl^{2-}})$ which is used to non-dimensionalize the diffusion equation for $c_{Fl^{2-}}$.
    
   All concentrations are kept constant far from the electrode ($z=L$) at their respective initial values, while flux boundary conditions are employed at the electrode surface. In particular, the consumption flux of \ce{H+} is set by the Faradaic current density $i_f$ as
    $\pdv{c_{\ce{H+}}^{*}}{z^{*}} = \frac{-\abs{i_f}L}{FD_{H^+}c_{H^+}^0}$, with  $F$ denoting the Faraday constant, while all other fluxes are zero at $z =0$.
    It is important to note that $i_f$ is not the applied current density $i$, but has been modified to take the contribution of a capacitive current into account. This is achieved by using a constant capacitance similarly as done by \citeauthor{Bonnefont2001}\cite{Bonnefont2001}. Assuming a Stern layer thickness $\lambda_S =  1\si{\nm}$, a permittivity 10 times the vaccum permittivity \cite{Guidelli1992} $\epsilon_0$, and that the whole potential drop occurs within the Stern layer, the capacitance $C$ is estimated  to be about $C \approx$ 88 $\mu\text{F}/\text{cm}^2$. The double layer capacitance of platinum (measured in 0.1 M \ce{KClO4}) was found to be in a similar range (20 - 120 $\mu\text{F}/\text{cm}^2$) \cite{Tymosiak-Zielinska2001}. $i_f$ is then related to the measured time-dependent potential ($\phi$) change by $i_f = i-C\frac{d\phi}{dt}$. Since the exact value of C in our measurements is not known, results for $0 \leq C \leq 120~ \mu\text{F}/\text{cm}^2$ have also been presented in Figure 3c in the main text.

\begin{table}
\resizebox{\columnwidth}{!}{
  \caption{Parameters and associated used in the model}
  \label{tbl:paramVal}
  \begin{tabular}{|c|c||c|c|}
    \hline
    Parameter  & Value (units) & Parameter & Value (units) \\
    \hline
    $D_{H^+}$ & $9.3\times 10^{-9}$ ($m^2/s$)\cite{Rudd} & T & $8\times 10^{-6}$ ($M$) \\
    \hline
    $D_{OH^-}$ & $4.62\times 10^{-9}$  ($m^2/s$)\cite{Rudd} & $k_{f,{Fl}}$ & $k_f$\\
    \hline
    $D_{Fl^{2-}}$ & $0.42\times 10^{-9}$ ($m^2/s$)\cite{Casalini2011} & $K_{f_{eq}}$ & $4.36\times10^{-7}$ ($M^{-1}$)\cite{Diehl1989b} \\
    \hline
    $k_f$ & $1.4\times10^{11}$ ($M^{-1}s^{-1}$)\cite{Paldus1976b} & Area  & $4\pi\times 10^{-4}$ ($m^2$)\\
    \hline
    $k_b$ & $2.6\times10^{-5}$ ($s^{-1}$)\cite{Paldus1976b} & $C_S$ = $10\epsilon_0/\lambda_s$ & 88 ($\mu \text{F}/\text{cm}^2$)\cite{Bonnefont2001} \\
    \hline
  \end{tabular}
  }
\end{table}

The large reaction rate constants and the associated large Damköhler numbers ($Da \gtrapprox 10^{6}$) render the system of equations very stiff. 
To nonetheless numerically handle them efficiently, an implicit integrating factor formulation was adopted \cite{Nie2006,Chou2007}.

A second order central difference scheme is used to discretize spatial gradients. The numerical domain of length $L$ is divided into $N+1$ grid points such that $\Delta z =\frac{L}{N}$. We then obtain a set of equations of the form:
 \begin{equation}
 \begin{split}
        \pdv{\bm{c^*}_k}{t^*}= \frac{D_{r,k}}{\Delta z^2}&\left(\begin{bmatrix}-2&2&0\cdots \\1&-2&1&\cdots \\\vdots & \vdots & \ddots & \vdots\\0&\cdots &1&-2\end{bmatrix}~\begin{bmatrix}c^*_k(1)\\\vdots\\c^*_k(N) \\ \end{bmatrix}\right.
    \\&\left.+ ~\begin{bmatrix}\pm J_k\Delta z\\0\\\vdots \\~\\~\\c^*_k(N+1) \end{bmatrix}\right)\pm Da\times f(\bm{c}^*)
\end{split}
\end{equation}
Here $\bm{c^*_k} = c_k(m)$ is the (non-dimensional) spatially discretized concentration ($1\leq m\leq N+1$, but continuous in time) of ions and $J_k$ is the constant flux of the ions at the electrode surface. In our simulations:
\begin{subequations}
\begin{align}
    c^*_{H^+}(N+1) & = 1\\
    c^*_{OH^-}(N+1) & = 1\\
    c^*_{Fl^{2-}}(N+1) & = \frac{c_{Fl^{2-}}}{T}
\end{align}
\end{subequations}
and,
\begin{subequations}
\begin{align}
\begin{split}
    J_{H^+} &=\\&
    \begin{cases}
      0, &\text{if}\ \abs{i_{Cap}}>\abs{i}\\
      \frac{\left(-\abs{i}+\abs{i_{Cap}}\right)L}{FD_{H^+}c_{H^+}^0} ,&\text{otherwise}
    \end{cases}
\end{split}\\
    J_{OH^-} &= 0\\
    J_{Fl^{2-}} &= 0
\end{align}
\end{subequations}
where $i_{Cap}$ is the time dependent capacitive current density as described in the main text.

The above set of equations are stiff due to large reaction rate constants (and related $Da$) for the non-linear reaction terms $f(c)$. Using an implicit scheme such as Backward-Time-Central-Space (BTCS), would therefore require very small time steps and consequently a large run-time. We instead integrate in time according to the implicit integrating factor (IIF) scheme presented in \citeauthor{Nie2006} \cite{Nie2006}. A second order approximation of the $f(c)$-term is used, while employing the trapezoid rule to approximate the integration of the time dependent $i_{Cap}$. The set of non-linear equations are then solved at each time step using the \emph{fsolve} function in Matlab (R2019b). The numerical scheme was validated, first with the analytical solution of the linear reaction-diffusion equation used in \citeauthor{Nie2006} \cite{Nie2006}. We further compared the results of the IIF with the BTCS scheme (for our system) for the highest current density used in this work ($\abs{i}$ = 5.59 $\mu \text{A}/\text{cm}^2$). The results compare well as shown in Figure \ref{fig:validation}, confirming the proper implementation of the numerical scheme.

\subsection{Repeat experiments}
\label{repeats}
Figure\ref{repeats} shows the measured pH profiles in a repeated experiment has been provided to highlight the reliability of the measurement method. A comparison of the pH front with the model results is also shown. Similar to the results presented in the main text, the experimental profiles are in good agreement at the two lower current densities and deviate from the 1D diffusion model at the two highest current densities. 
\begin{figure}
\centering
\includegraphics[width=\columnwidth]{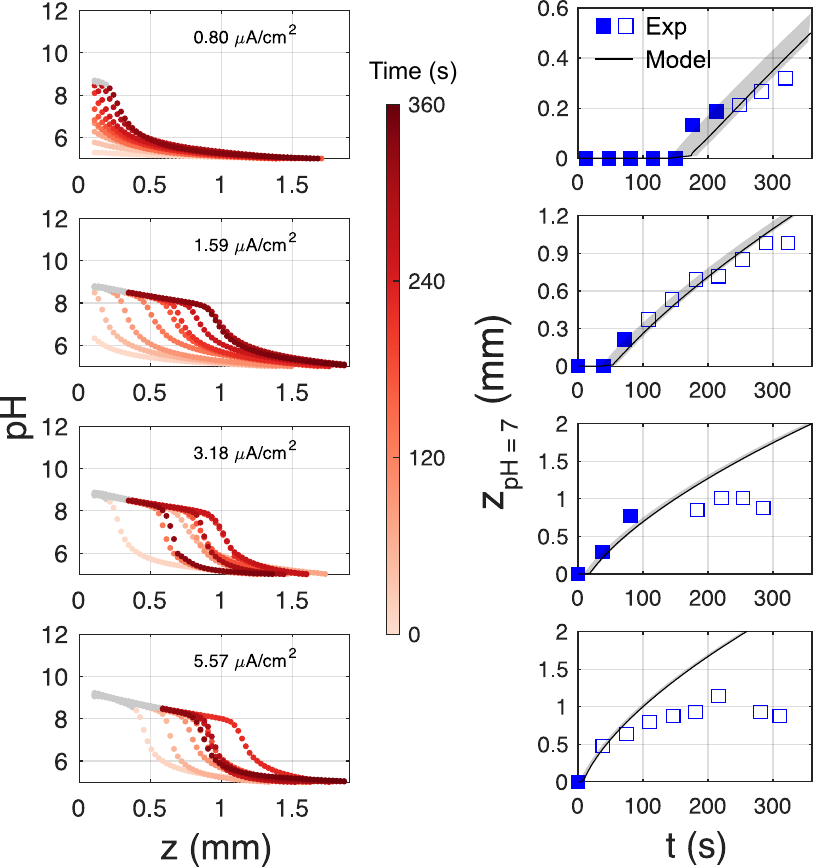}
    \caption{Experimental repeats for all $\abs{i}$ presented in the main text. Left: pH profile measured. pH $>$ 8.5 has been grayed out due to the uncertainty in measurement described in the main text. Right: Comparison of depletion length, $z_{pH = 7}$, of experiment versus model. The shaded region indicates the model results over a range of capacitance $0 \leq C \leq 120~ \mu\text{F}/\text{cm}^2$ (solid line with C = $88~ \mu\text{F}/\text{cm}^2$). Open squares show pH front after the first appearance of inhomogeneity described in the main text.}
    \label{fig:repeat}
\end{figure}

\subsection{Sulfates comparison}
\label{sulfateSalts}

To account for the buffering capacity of sulfates ($\ce{Na2SO4}/\ce{H2SO4}$) we adjusted the model to include the following reaction couples:
\begin{align}
    \ce{  H+ + SO4^{2-} &<=>[k_{b,S}][k_{f,S}] HSO4-}\\
    \ce{  Na+ + SO4^{2-} &<=>[k_{b,Na}][k_{f,Na}] NaSO4-}
\end{align}
There is an extra bulk reaction term due to \ce{HSO4-} ionization in the \ce{H+} equation. Furthermore, additional reaction diffusion equations (\ce{SO4^{2-}} and \ce{NaSO4-} here) and have to be taken into account. The initial concentrations of the ions are estimated based on the total dissolved \ce{Na2SO4} and pH of the solution. The dissociation constants of \ce{HSO4}\cite{Hamer1934,Covington1964a,Wu1995a} and \ce{NaSO4-}\cite{Pytkowicz1969,Santos1975a} are taken from the literature ($K_{HSO4} = 0.0103$ and $K_{NaSO4} = 0.5$). Also for the forward rate constants $k_{f,Na} = k_{f} \times 0.1$ and $k_{f,S}$ \cite{Irish1970b} $= k_f\times 5$ are taken.

\begin{figure}
\centering
\includegraphics[width=\columnwidth]{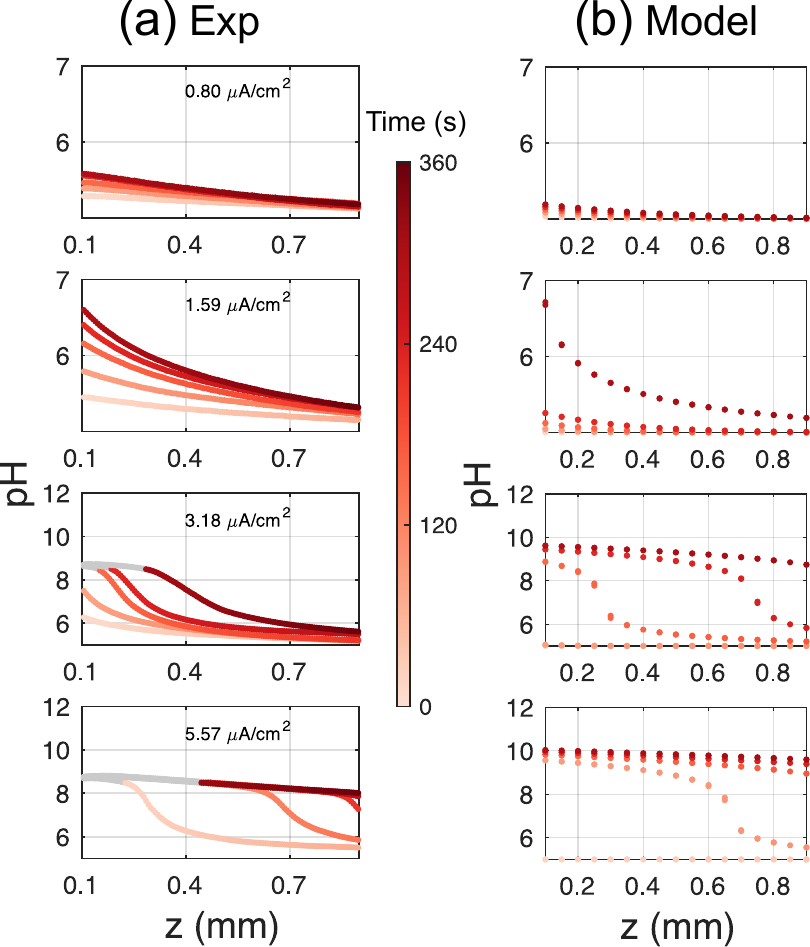}    
\caption{pH versus Depth for the case with sulfates: Experiments versus Simulations. (a) The experimental profiles for all $\abs{i}$ values. (b) Numerical pH profiles based on values discussed in the text.}
\label{fig:sulphatesDepth}
\end{figure}

Figure \ref{fig:sulphatesDepth} shows a comparison of the full pH profiles for the 4 different current densities considered. The buffer effect at the two lower current densities is captured well by the model, although the profiles do not match exactly at all times. Finally, the two highest current densities have been presented here only for the sake of completeness, as (discussed in the main text) experiments at these values of $\abs{i}$ are marked by the appearance of inhomogeneous fluorescein intensity in a plane and therefore do not follow a 1D diffusion approximation.
\FloatBarrier
\FloatBarrier

\bibliography{bib/Main_Submission_Final_proofs}